\PassOptionsToPackage{table}{xcolor}
\documentclass[sigconf]{acmart}
\setkeys{acmart.cls}{balance=false}

\usepackage{amsmath}

\usepackage{amssymb}
\usepackage{algorithm}
\usepackage{algorithmic}
\usepackage{subcaption}
\usepackage{multirow}
\usepackage{xspace}
\usepackage{enumitem}
\usepackage{framed}
\definecolor{shadecolor}{rgb}{1.0,0.98,0.88}% warm yellow matching original

\newcommand{\methodname}{PRISM\xspace}
\newcommand{\cmark}{\textcolor{teal}{\textbf{\checkmark}}}
\newcommand{\xmark}{\textcolor{red!70!black}{\textbf{$\times$}}}

\AtBeginDocument{%
  \tolerance=10000
  \hbadness=10000
  \vbadness=10000
  \providecommand\BibTeX{{%
    \normalfont B\kern-0.5em{\scshape i\kern-0.25em b}\kern-0.8em\TeX}}}

\copyrightyear{2026}
\acmYear{2026}
\setcopyright{cc}
\setcctype{by}
\acmConference[CIKM '26]{Proceedings of the 35th ACM International Conference on Information and Knowledge Management}{November 07--11, 2026}{Rome, Italy}
\acmBooktitle{Proceedings of the 35th ACM International Conference on Information and Knowledge Management (CIKM '26), November 07--11, 2026, Rome, Italy}
\acmDOI{10.1145/3799682.3840694}
\acmISBN{979-8-4007-2539-5/2026/11}
\begin{document}
\renewcommand{\shortauthors}{A. A. Shukla et al.}

\title{Prototype-Rectified Iterative Self-supervised Manifold Denoising under Severe Acoustic Shift}

\settopmatter{authorsperrow=4}

\author{Ashish Anand Shukla}
\email{shukla24@iiserb.ac.in}
\affiliation{%
  \institution{Indian Institute of Science Education and Research}
  \city{Bhopal}
  \country{India}
}

\author{Rini Smita Thakur}
\email{rinithakur@iiserb.ac.in}
\affiliation{%
  \institution{Indian Institute of Science Education and Research}
  \city{Bhopal}
  \country{India}
}

\author{Aryan Das}
\email{aryan.das2021@vitbhopal.ac.in}
\affiliation{%
  \institution{Vellore Institute of Technology}
  \city{Bhopal}
  \country{India}
}

\author{Vinod K. Kurmi}
\email{vinodkk@iiserb.ac.in}
\affiliation{%
  \institution{Indian Institute of Science Education and Research}
  \city{Bhopal}
  \country{India}
}

\begin{abstract}
Audio-Text Foundation Models~(ATMs) fail catastrophically under severe acoustic noise, yet existing adaptation strategies either rely on gradient-based Test-Time Adaptation~(TTA), which reinforces noise rather than signal, or on prompt tuning that requires privileged noise annotations unavailable at inference. We address these failures with \textit{\methodname{} (Prototype-Rectified Iterative Self-supervised Manifold Denoising)}, a training-free, source-free TTA framework grounded in the \textit{Affine Noise Hypothesis}: severe acoustic noise induces a low-rank affine shift in the multimodal latent space, with more than 90\% of distortion energy confined to the leading 60 principal components. \methodname{} estimates and reverses this distortion from an unlabeled target batch using frozen text prototypes as geometric anchors via three closed-form geometric corrections compiled into a single static projection matrix by Affine Bias Regression. In inference, adaptation reduces to one matrix-vector multiplication in $9 \times 10^{-4}$ ms, making it substantially faster than gradient-based TTA while requiring no additional training. On UrbanSound8K, \methodname{} improves over the zero-shot baseline by $+12.94$\,pp and surpasses an oracle-assisted TTA baseline by $+9.41$\,pp, which requires privileged augmented noise prompts \methodname{} never sees. We further identify the \textit{Polyphonic Trap}, a principled failure mode of subspace deflation for broadband classes, and resolve it via Confidence-Aware Regression~(CAR), recovering up to 8.16\,pp for the worst-affected class. The code is available at \href{https://github.com/Ashish-1108/PRISM}{https://github.com/Ashish-1108/PRISM}.
\end{abstract}



\begin{teaserfigure}
\centering
\begin{minipage}[t]{0.52\textwidth}
    \vspace{0pt}
    \centering
    \includegraphics[width=\linewidth]{figures/teaser.pdf}
\end{minipage}
\hfill
\begin{minipage}[t]{0.45\textwidth}
    \vspace{0pt}
    \centering
    \includegraphics[width=1\linewidth]{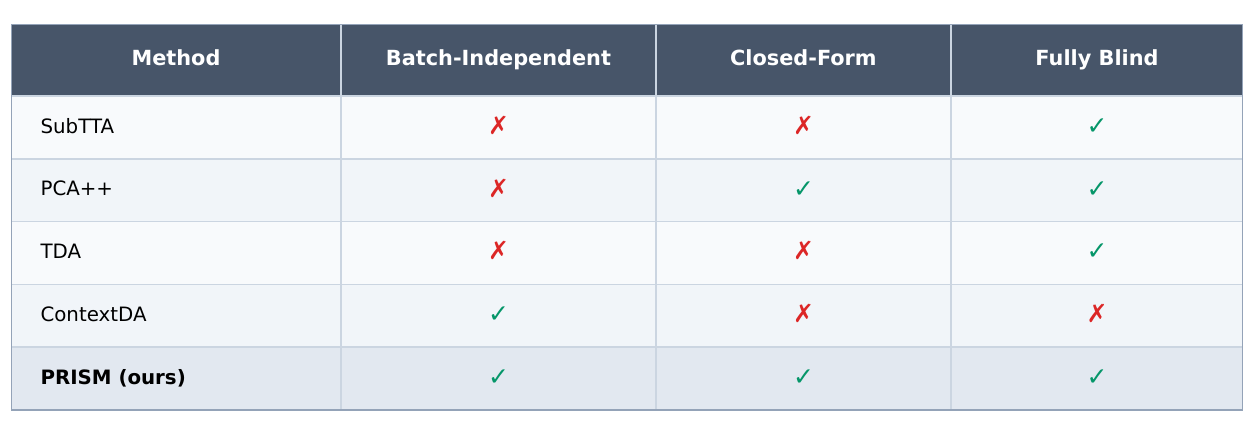}
    \includegraphics[width=0.68\linewidth]{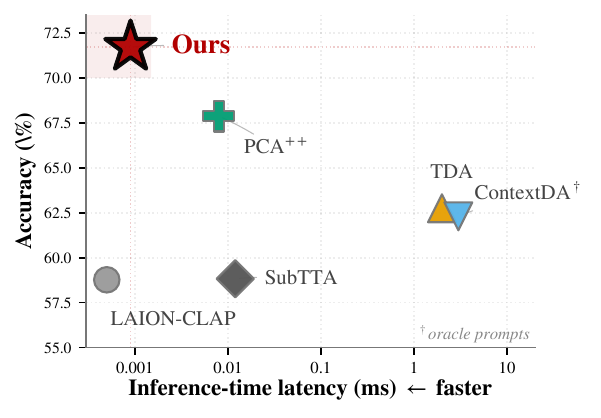}
\end{minipage}
\caption{
\textbf{Overview of \methodname{}.}
(a) ContextDA (top-left) requires privileged oracle prompts, whereas \methodname{} (bottom-left) compiles geometric corrections into a static projection matrix $\mathbf{W}$ for fully blind adaptation.
(b) Top-right: \methodname{} satisfies all key desiderata for edge inference.
(c) Bottom-right: \methodname{} achieves higher accuracy with substantially lower inference latency than the strongest-performing baseline.
}
\label{fig:teaser_master}
\Description{Overview figure. Left: pipeline comparison showing ContextDA needing oracle noise prompts vs. PRISM doing blind adaptation. Top right: a table comparing key properties. Bottom right: scatter plot of accuracy vs. inference latency, showing PRISM is both faster and more accurate.}
\end{teaserfigure}
\maketitle

%%%%%%%%%%%%%%%%%%%%%%%%%%%%%%%%%%%%%%%%%%%%%%%%%%
\section{Introduction}
\label{sec:intro}

The rapid adoption of ATMs in real-world applications has outpaced their robustness to uncontrolled acoustic environments. While models like CLAP and its latest variants ~\cite{wu2023laionclap, elizalde2023msclap, paissan24_interspeech, niizumi2025m2d, ghosh2025reclap} demonstrate strong zero-shot generalization in standardized benchmarks, they are increasingly integrated into field operations such as fiber-optic acoustic monitoring ~\cite{sun2025clap}, bio-acoustic animal recognition ~\cite{miao2025multi}, and underwater vessel classification ~\cite{li2024clapp}. In these inference scenarios, operating conditions are routinely hostile: signal-to-noise ratios frequently drop below 0 dB, and target events are submerged beneath non-stationary interference, dense polyphony, and heavy reverberation~\cite{zhi2024confusing, yang2026clapmamba}. When running on resource-constrained edge hardware, these systems cannot tolerate the computational overhead of cloud-based fallbacks or iterative optimization loops, driving a recent shift toward backpropagation-free frameworks~\cite{ebats2025}.

\begin{shaded*}
\centering
\itshape\bfseries
Can we design a TTA method for ATMs that autonomously restores semantic alignment under severe acoustic shift, without gradients, source data, or augmented noise prompts, while remaining fast enough?
\end{shaded*}

% 2. Current existing solutions and what they lack
\textit{A proper answer has yet to emerge, and existing adaptation strategies introduce new failure modes when confronted with environmental corruption.} Current approaches can be grouped into three paradigms, all of which are fundamentally mismatched to harsh acoustic inference. Static geometric alignment methods apply fixed projections or source-trained mappings to bridge the modality gap~\cite{deshmukh2024domain}, but each target environment induces a unique rotation, translation, or low-rank distortion, making one-shot global corrections too rigid. Gradient-based TTA frameworks optimize logit distributions or entropy at inference time~\cite{wang2021tent, shi2025emotta}, and generative spoken language models adapt via dynamic prompt interleaving~\cite{slm_tta2025}. Yet under ambiguous polyphonic noise they suffer from severe confirmation bias: incorrect high-confidence pseudo-labels are fed back into the adaptation loop, effectively training the model on the noise floor rather than the foreground signal~\cite{farina2024frustratingly}. Prompt-based optimization ~\cite{shu2022tpt, liang2023adapting} dynamically tunes textual embeddings to mirror corrupted acoustics, but it relies on privileged noise-type annotations that are unavailable at inference, and introduces iterative backpropagation overhead that is incompatible with real-time audio streams. Taken together, prior work lacks a lightweight, training-free mechanism that can isolate and invert environment-specific geometric corruption without reinforcing noise or demanding prohibitive compute.

% 3. Our motivation and solution to the problem raised
\textit{We argue that the bottleneck is not classifier mismatch but the geometric corruption of the multimodal latent space itself, a corruption that can be modeled, measured, and reversed analytically.} We formalize the \textit{Affine Noise Hypothesis}, where severe additive acoustic noise induces an approximately low-rank affine distortion in the latent space, with empirical Singular Value Decomposition (SVD) confirming that over $90$\% of the distortion energy concentrates in principal dimensions. This insight renders full gradient-based adaptation unnecessary. Instead, we propose \textbf{Prototype-Rectified Iterative Self-supervised Manifold Denoising (\methodname{})}, a training-free, source-free, and noise-prompt-free transductive TTA framework that operates entirely in the embedding space. \methodname{} estimates the target-domain distortion from an unlabeled calibration batch using frozen text prototypes as clean semantic anchors. It applies three closed-form geometric corrections---Orthogonal Procrustes Cross-modal Alignment~(OPCA), Class-Conditioned Variance Deflation~(CCVD), and Per-Class Residual Translation~(PCT)---and compiles them into a single static projection matrix using Affine Bias Regression~(ABR). Calibration is transductive, but inference is strictly batch-independent: once compiled, each future sample is adapted through a single matrix-vector multiplication in $9 \times 10^{-4}$ ms, which is orders of magnitude faster than gradient-based TTA with zero trainable parameters. We further identify the \textit{Polyphonic Trap}, a principled failure mode where broadband sound classes geometrically overlap with noise directions during subspace deflation, and resolve it via Confidence-Aware Regression~(CAR), recovering up to $8.16$\,pp for the worst-affected class. Our contributions are as follows:
\begin{itemize}[leftmargin=15pt, topsep=2pt, itemsep=1pt, parsep=0pt]
    \item \textbf{\textit{Affine Noise Hypothesis.}} We formulate severe acoustic shift as a low-rank affine distortion of the multimodal latent space, providing a theoretical basis for reversing the modality gap without paired clean-noisy audio, target labels, or augmented noise prompts.
    \item \textbf{\textit{\methodname{} Framework.}} We introduce a source-free, training-free TTA pipeline that restores audio-text alignment through three closed-form geometric operations embedded in a self-supervised transductive calibration loop, cleanly separating calibration from strictly batch-independent inference.
    \item \textbf{\textit{Polyphonic Trap and CAR.}} We formalize a principled limitation of subspace deflation for broadband classes and resolve it with CAR. By bridging the difference between standard \methodname{} and \methodname{}+CAR, we achieve an $+8.16$\,pp improvement for the most severely degraded class, successfully recovering a substantial portion of the performance gap relative to the ideal zero-shot baseline.
    \item \textbf{\textit{Robustness and Edge Efficiency.}} \methodname{} achieves state-of-the-art accuracy on both additive-noise benchmarks (US8K, ESC-50) while operating at sub-millisecond inference latency with zero gradient updates, making foundation model inference viable on resource-constrained edge hardware.
\end{itemize}

%%%%%%%%%%%%%%%%%%%%%%%%%%%%%%%%%%%%%%%%%%%%%%%%%%
\section{Related Work}

\noindent\textbf{Audio-Text Foundation Models and the Modality Gap:}
CLAP models \cite{wu2023laionclap,elizalde2023msclap} use dual encoders trained via symmetric cross-entropy to project audio and text into a shared latent space. While highly effective, contrastive objectives intrinsically generate a geometric modality gap \cite{deshmukh2024domain,das2026nexus,das2026hypercap,das2026uavls,DAS2025113152,das2026unityattentionflownetworks,das_lts}, i.e., a systematic spatial separation between the audio and text manifolds due to the disparate statistical densities of the underlying modalities. In clean acoustic scenarios, relative class geometry remains intact, permitting successful zero-shot matching. However, when deployed in environments with severe additive noise, the corrupted audio embeddings are pulled into out-of-distribution sub-manifolds dominated by the background interference, breaking the relative geometry and causing zero-shot retrieval to fail. Static projection methods, such as covariance whitening or linear probing, fail to resolve this because they treat noise as a rigid, monolithic shift rather than an instance-specific geometric distortion.

\noindent\textbf{Test-Time Adaptation (TTA) and Transductive Learning:}
Addressing severe domain shifts without source data or gradient-heavy fine-tuning, the field has converged on transductive and test-time adaptation (TTA) frameworks. Traditional machine learning relies on inductive reasoning, whereas transductive algorithms aim to make accurate predictions specifically for a given, observed set of unlabeled test data.
Methods such as TENT \cite{wang2021tent} minimize the Shannon entropy of output logits, implicitly performing unsupervised clustering at test time. However, when applied to audio under extreme degradation ($<0$ dB), entropy minimization forces high-confidence predictions on ambiguous signals, leading to confirmation bias where the model collapses into predicting the noise floor. Single-Utterance Test-Time Adaptation (SUTA) \cite{suta2022} and its modern dynamic model-bank extensions \cite{dmsuta2025} attempt to adapt based on confidence heuristics over short utterances, but similarly suffer from confirmation bias when noise entirely dominates the utterance.

Expectation-Maximization TTA (Emo-TTA) \cite{shi2025emotta} estimates empirical class distributions using Bayesian text priors to compute soft assignments, effectively anchoring the adaptation process against noise collapse. Test-Time Prompt Tuning (TPT) \cite{shu2022tpt} and Contrastive Domain Vector Optimization \cite{deshmukh2024domain} dynamically modulate the textual input space via gradient descent. They generate augmented views of the noisy audio and tune continuous prompt tokens to mirror the corrupted audio geometry via self-entropy loss. Similarly, Confidence-Enhanced Adaptation (CEA) \cite{cea2024} performs frame-level sequential adaptation utilizing a confidence-aware weight scheme combined with Short-Term Consistency Regularization to preserve semantic temporal continuity in speech. 

While these methods achieve reliable alignment and avoid pseudo-labeling error accumulation (as seen in CoNMix \cite{conmix2023}), their reliance on inference-time backpropagation introduces latency bottlenecks incompatible with real-time audio streams. Recent work such as E-BATS \cite{ebats2025} explicitly attempts to bypass this via backpropagation-free prompt adaptation, highlighting the critical need for edge-efficient solutions. In contrast, \methodname abandons explicit logit-level manipulation \cite{cea2024, 10944001}, prompt adaptation, and gradient descent entirely. It operates strictly via closed-form linear algebra in the embedding space, delivering robust transductive adaptation at sub-millisecond speeds.

%%%%%%%%%%%%%%%%%%%%%%%%%%%%%%%%%%%%%%%%%%%%%%%%%%
\section{Theoretical Foundations}

%%%%%%%%%%%%%%%%%%%%%%%%%%%%%%%
\begin{figure}[t]
  \centering
  \includegraphics[width=0.60\linewidth]{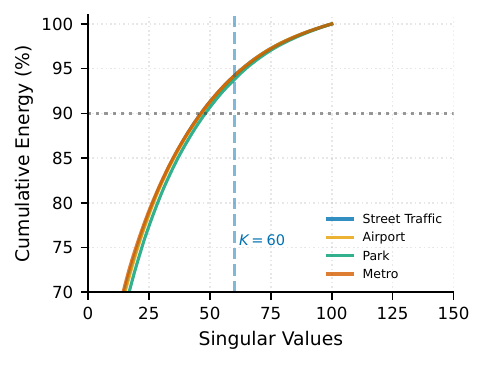}
  \caption{
  Empirical validation of the Affine Noise Hypothesis: Cumulative Singular Value energy of noise-induced distortion across four environments. Dashed lines denote retained rank \(K\), while horizontal guides indicate 90\% energy.
  }
  \label{fig:spectral_decay}
  \Description{Four cumulative energy curves, one per noise background. Each rises steeply and crosses the 90\% threshold near rank 60, supporting the idea that noise distortion lives in a small low-rank subspace.}
\end{figure}
%%%%%%%%%%%%%%%%%%%%%%%%%%%%%%%

%%%%%%%%%%%%%%%%%%%%%%%%%%%%%%%%%%%%%%%%%%%%%%%%%%
\subsection{Preliminary}
Let \(f_A : \mathcal{X} \rightarrow \mathbb{R}^d\) and \(f_T : \mathcal{Y} \rightarrow \mathbb{R}^d\) represent the frozen audio and text encoders of a contrastive foundation model, where \(x \in \mathcal{X}\) is a clean audio sample and \(y \in \mathcal{Y}\) is its textual descriptor. The embeddings \(e = f_A(x)\) and \(t = f_T(y)\) reside in a joint \(d\)-dimensional space. In high-SNR environments, the zero-shot classifier accurately identifies the correct class \(c\) via \(\arg\max_{c} \langle e, t_c \rangle\).
When subjected to additive environmental noise $\delta$, the received signal becomes $x' = x + \delta$. Under severe interference, the non-linear transformation through $f_A$ produces a heavily corrupted embedding $e' = f_A(x')$. Because the acoustic energy of $\delta$ dominates $x$, $e'$ drifts drastically toward the latent representation of the noise concept, such that $\langle e', t_c \rangle < \langle e', t_{noise} \rangle$.

%%%%%%%%%%%%%%%%%%%%%%%%%%%%%%%%%%%%%%%%%%%%%%%%%%
\subsection{ Affine Noise Hypothesis}\label{sec:anh}
We postulate the \textit{Affine Noise Hypothesis}: Although the raw waveform corruption $x' = x + \delta$ undergoes a complex non-linear transformation via the transformer encoder, the resulting distortion in the latent space manifests primarily as an affine shift constrained within a low-rank subspace. 
Mathematically, the corrupted embedding $e'$ can be modeled as:
\begin{equation}
e' \approx R e + \Delta_{noise}
\label{eq:anh}
\end{equation}
where $R \in \mathbb{R}^{d \times d}$ is an orthogonal rotation matrix reflecting the skew induced by the noise, and $\Delta_{noise} \in \mathbb{R}^d$ is a translation vector representing the dense acoustic centre of the background. 

If this hypothesis holds, finetuning or prompt optimization becomes unnecessary and may even move the model away from the optimal solution. Instead, the modality gap can be corrected through an unsupervised affine inverse transformation in the embedding space. To preliminarily validate this hypothesis, we examine the singular value decay of the noise-induced distortion across four acoustically diverse environments. For each environment, we subtract every clip's clean embedding from its noisy counterpart, collect the resulting residuals into a matrix, and compute its SVD; this paired comparison is purely diagnostic and is not part of the \methodname{} pipeline. As shown in Figure~\ref{fig:spectral_decay}, the singular value curves for all four environments collapse onto the same profile: near-maximal energy in components 1--5, then an abrupt ``elbow,'' followed by a long flat tail near zero. This confirms that the noise distortion $\Delta_{noise}$ is not isotropic, but instead has a strongly concentrated low-rank structure. Because $>$90\% of the energy is concentrated in a low-rank subspace, the signal subspace is largely preserved. This observation supports our view that severe acoustic noise induces a structured geometric distortion that can be separated from the semantic signal.
We further validate the hypothesis through our results and analysis in \S~\ref{sec:results}

%%%%%%%%%%%%%%%%%%%%%%%%%%%%%%%%%%%%%%%%%%%%%%%%%%
\section{Proposed Approach}
\begin{figure*}[t]
\centering
\includegraphics[width=0.85\textwidth]{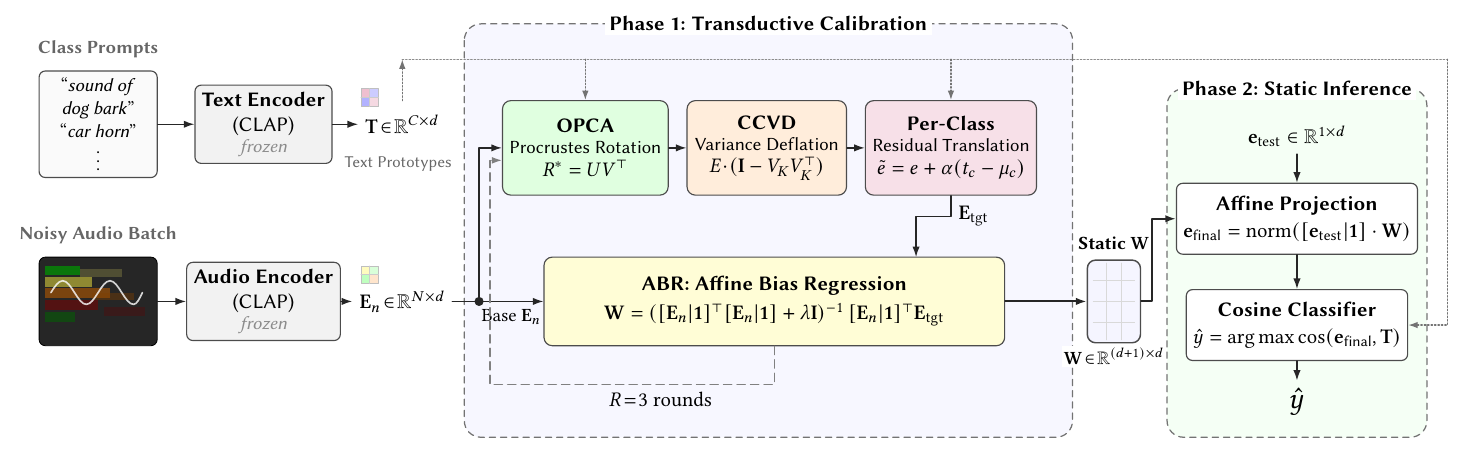}
\caption{Proposed Framework.
  \textbf{Phase~1:} Noisy audio embeddings~$E_n$ are corrected via three closed-form geometric operations (OPCA, CCVD, and Per-Class Residual Translation) using frozen text prototypes~$T$ as semantic anchors. Corrections are compiled into a static projection matrix~$W$ via Affine Bias Regression (ABR).
  \textbf{Phase~2:} The frozen~$W$ adapts each future sample via a single matrix multiplication without requiring gradients, batch statistics, or parameter updates.}
\label{fig:architecture}
\Description{Two-phase diagram. Left shows the calibration loop: noisy audio embeddings are corrected via OPCA, CCVD, and per-class translation, then distilled into a single projection matrix W. Right shows inference: a new sample is adapted by multiplying by W and classified against text prototypes.}
\end{figure*}

%%%%%%%%%%%%%%%%%%%%%%%%%%%%%%%

In this section, we detail the proposed \methodname{} framework, a source-free, noise-agnostic transductive method designed to reverse the structured affine perturbation identified by the Affine Noise Hypothesis (\S~\ref{sec:anh}) using the foundation model's frozen text prototypes as geometric anchors. The method operates in two stages: (i) a batch-dependent transductive calibration phase comprising three closed-form geometric operations iterated over $R$ rounds, and (ii) a strictly batch-independent inference phase. Let $E_n \in \mathbb{R}^{N \times d}$ denote the $\ell_2$-normalized noisy audio embeddings for $N$ test samples in a $d$-dimensional joint embedding space, and let $T \in \mathbb{R}^{C \times d}$ denote the matrix of $C$ frozen text prototypes, where $t_c$ is the $c^{\text{th}}$ prototype. Throughout, $\epsilon > 0$ denotes a small numerical stability constant.
The three stepwise operations are:
\begin{enumerate}[topsep=2pt, itemsep=1pt, parsep=0pt, leftmargin=*]
    \item \textbf{Orthogonal Procrustes Cross-modal Alignment (OPCA, \S~\ref{sec:opca})}: aligns the corrupted audio manifold onto the text prototype manifold via confidence-gated Orthogonal Procrustes rotation.
    \item \textbf{Class-Conditioned Variance Deflation  (CCVD, \S~\ref{sec:ccvd})}: identifies noise-dominant embedding directions through Fisher Linear Discriminant Analysis and removes them via orthogonal projection.
    \item \textbf{Per-Class Residual Translation (\S~\ref{sec:pct})}: translates each sample toward its predicted text prototype centroid to resolve residual inter-class displacement.
\end{enumerate}
After each round, the accumulated corrections are compiled into a single static \textbf{Affine Bias Regression (ABR)} matrix via ridge regression (\S~\ref{sec:abr}). The full iterative procedure is illustrated in Figure~\ref{fig:architecture} and detailed in \S~\ref{sec:calibration}.

%%%%%%%%%%%%%%%%%%%%%%%%%%%%%%%%%%%%%%%%%%%%%%%%%%
\subsection{Orthogonal Procrustes Cross-modal Alignment (OPCA)}
\label{sec:opca}

Acoustic noise induces a systematic rotation of the audio embedding manifold away from the text prototype manifold on the unit hypersphere $\mathbb{S}^{d-1} = \{x \in \mathbb{R}^d : \|x\|_2 = 1\}$. OPCA estimates and corrects this rotation via the Orthogonal Procrustes problem~\cite{schonemann1966generalized}, using confidence-gated pseudo-labels to suppress the influence of unreliable samples, following the broader motivation of uncertainty-aware pseudo-label refinement in source-free adaptation ~\cite{10943558, thakur2025gradcl}.

\paragraph{Pseudo-Labelling and Confidence Gating.}
Given the current embeddings $E_{\text{cur}}$ (initialised to $E_n$ at round $r{=}1$), cosine similarity logits are computed as $L = E_{\text{cur}}\, T^\top \in \mathbb{R}^{N \times C}$. The predicted class and per-sample confidence for sample $i$ are:
\begin{equation}
\hat{y}_i = \arg\max_c L_{ic}, \qquad \kappa_i = \max_c L_{ic}.
\label{eq:pseudo}
\end{equation}
Let $\tau = Q_{1-p}(\kappa)$ denote the $(1{-}p)$-th quantile of the confidence distribution. The confident set $\mathcal{C} = \{i : \kappa_i \geq \tau\}$ retains the top-$p$ fraction of samples (default $p = 0.8$).

\paragraph{Confidence-Weighted Class Centroids.}
For each class $c$ with at least one confident member, the audio centroid is:
\begin{equation}
M_n^c = \sum_{\substack{i \in \mathcal{C},\; \hat{y}_i = c}} \frac{\kappa_i}{\sum_{\substack{j \in \mathcal{C},\; \hat{y}_j = c}} \kappa_j} \; e_i,
\label{eq:centroid}
\end{equation}
where $e_i$ denotes the $i$-th row of $E_{\text{cur}}$. The text centroid for class $c$ is simply $M_t^c = t_c$. Let $\mathcal{V}$ denote the set of classes with at least one confident member, and let $\mu_n$, $\mu_t$ be the arithmetic means of $\{M_n^c\}_{c \in \mathcal{V}}$ and $\{M_t^c\}_{c \in \mathcal{V}}$, respectively.

\paragraph{Optimal Orthogonal Rotation.}
The cross-covariance matrix of mean-centred centroids is:
\begin{equation}
H = \bigl(M_n^{\mathcal{V}} - \mu_n\bigr)^\top \bigl(M_t^{\mathcal{V}} - \mu_t\bigr) \in \mathbb{R}^{d \times d}.
\label{eq:crosscov}
\end{equation}
Its Singular Value Decomposition (SVD), $H = U \Sigma V^\top$, yields the optimal rotation $R^* = U V^\top$, which minimises $\sum_{c \in \mathcal{V}} \|M_n^c R - M_t^c\|_2^2$. Crucially, since the Affine Noise Hypothesis (Eq.~\ref{eq:anh}) posits that the forward noise process applies a rotation $R$ to clean embeddings ($e' \approx Re + \Delta_{\text{noise}}$), recovery requires the \emph{inverse} rotation $R^{-1}$. For orthogonal $R$, $R^{-1} = R^\top$. The Procrustes solution $R^*$ directly estimates this inverse by finding the orthogonal map from the noisy audio manifold \emph{back} to the clean text-prototype manifold.

\paragraph{Adaptive Scale and Transformation.}
An adaptive isotropic scale factor prevents over-compression of the rotated manifold:
\begin{equation}
s = \mathrm{clamp}\!\left(\frac{\|M_t^{\mathcal{V}} - \mu_t\|_F}{\|M_n^{\mathcal{V}} - \mu_n\|_F + \epsilon},\; 0.8,\; 1.2\right).
\label{eq:scale}
\end{equation}
The OPCA-corrected embeddings are:
\begin{equation}
E' = \mathrm{normalize}\!\bigl(s \cdot (E_{\text{cur}} - \mu_n)\, R^* + \mu_t\bigr).
\label{eq:opca}
\end{equation}

\noindent\textit{Remark (Over-Rotation Phenomenon).}
A single Procrustes iteration is used intentionally. Additional iterations produce an over-rotation artefact: confident centroids collapse prematurely toward text prototypes, destroying the within-class geometry that CCVD relies upon. Empirically, multiple iterations degrade accuracy by an average of $-1.24$\,pp across environments.

%%%%%%%%%%%%%%%%%%%%%%%%%%%%%%%%%%%%%%%%%%%%%%%%%%
\subsection{Class-Conditioned Variance Deflation  (CCVD)}
\label{sec:ccvd}

While OPCA addresses global misalignment, intra-class variance driven by noise components remains. CCVD identifies these noise-dominant directions through Fisher Linear Discriminant Analysis~\cite{zhao2024linear} and removes them via orthogonal projection.

\paragraph{Confident Subset Selection.}
Pseudo-labels $\hat{l}$ and confidences $\kappa$ are recomputed from $E'$. Samples whose confidence exceeds the $(1{-}q)$-th empirical percentile form the filtered set $\mathcal{F}$ (default $q = 0.7$), with embeddings $E_{\mathcal{F}}$ and labels $\hat{l}_{\mathcal{F}}$.

\paragraph{Within-Class and Between-Class Scatter.}
Let $\mu = \frac{1}{|\mathcal{F}|} \sum_{i \in \mathcal{F}} e_i'$ be the global mean and $\mu_c = \frac{1}{n_c} \sum_{i \in \mathcal{F},\, \hat{l}_i = c} e_i'$ the per-class mean, where $e_i'$ is the $i$-th row of $E'$ and $n_c = |\{i \in \mathcal{F} : \hat{l}_i = c\}|$. The scatter matrices are:
\begin{align}
S_W &= \sum_{c=1}^{C} \sum_{\substack{i \in \mathcal{F} \\ \hat{l}_i = c}} (e_i' - \mu_c)(e_i' - \mu_c)^\top, \label{eq:sw} \\
S_B &= \sum_{c=1}^{C} n_c\, (\mu_c - \mu)(\mu_c - \mu)^\top. \label{eq:sb}
\end{align}

\paragraph{Noise Subspace Identification.}
Directions with high within-class variance relative to between-class variance characterise noise. They are the leading eigenvectors of the regularised Fisher ratio matrix:
\begin{equation}
F = \bigl(S_B + \epsilon\, I_d\bigr)^{-1} S_W,
\label{eq:fisher}
\end{equation}
where $\epsilon = 10^{-4}$ ensures numerical stability. Let $V_K \in \mathbb{R}^{d \times K}$ collect the $K$ eigenvectors of $F$ corresponding to the $K$ largest eigenvalues. These span the noise subspace.

\paragraph{Orthogonal Projection.}
The noise-decimated embeddings are:
\begin{equation}
E_{\text{ccvd}} = \mathrm{normalize}\!\bigl(E' \cdot (I_d - V_K V_K^\top)\bigr).
\label{eq:ccvd}
\end{equation}
The complement projector $P = I_d - V_K V_K^\top$ retains all signal in the orthogonal complement of the noise subspace.

%%%%%%%%%%%%%%%%%%%%%%%%%%%%%%%%%%%%%%%%%%%%%%%%%%
\subsection{Per-Class Residual Translation}
\label{sec:pct}

After OPCA and CCVD, a residual class-level displacement typically persists between audio cluster means and their corresponding text prototype vectors. We correct this via a soft per-class translation in the embedding space.

Let $\hat{y}_i = \arg\max_c (e_i^{\text{ccvd}} \cdot t_c)$ be the class prediction for sample $i$ under $E_{\text{ccvd}}$, and let $\mu_c^{\text{ccvd}} = \frac{1}{n_c} \sum_{j:\hat{y}_j = c} e_j^{\text{ccvd}}$ be the corresponding cluster mean. The translated embedding for sample $i$ is:
\begin{equation}
\tilde{e}_i = e_i^{\text{ccvd}} + \alpha \bigl(t_{\hat{y}_i} - \mu_{\hat{y}_i}^{\text{ccvd}}\bigr),
\label{eq:pct}
\end{equation}
with shift strength $\alpha \in (0, 1)$ (default $\alpha = 0.3$). The target embedding matrix is $E_{\text{tgt}} = \mathrm{normalize}(\tilde{E})$.

\noindent\textit{Remark (Embedding-Space vs.\ Logit-Space Corrections).}
The translation in Eq.~\ref{eq:pct} operates in the embedding space, not the logit space. Additive shifts applied to logits $L = E\,T^\top$ are annihilated under row-wise $\ell_2$ normalisation, because normalisation contracts all dimensions proportionally. Embedding-space translations do not suffer this degeneracy: the corrected direction toward the prototype is preserved after normalisation.

%%%%%%%%%%%%%%%%%%%%%%%%%%%%%%%%%%%%%%%%%%%%%%%%%%
\subsection{Affine Bias Regression (ABR): Compilation to a Static Map}
\label{sec:abr}

%%%%%%%%%%%%%%%%%%%%%%%%%%%%%%%%%%%%%%%%%%%%%
\begin{algorithm}[t]
\caption{\methodname: Prototype-Rectified Iterative Self-supervised Manifold Denoising}\label{alg:prism}
\small
\begin{algorithmic}[1]
\REQUIRE $E_n \in \mathbb{R}^{N \times d}$: $\ell_2$-normalised noisy audio embeddings;\;
         $T \in \mathbb{R}^{C \times d}$: $\ell_2$-normalised text prototypes;\;
         $R{=}3$: rounds;\; $p{=}0.8$: OPCA retention;\; $K{=}60$: CCVD dims;\;
         $q{=}0.7$: CCVD retention;\; $\alpha{=}0.3$: shift strength;\; $\lambda{=}0.01$: ridge coeff.
\ENSURE Denoised embeddings $E_{\text{cur}}$,\; static projection matrix $W_{\text{aug}}$
\STATE $E_{\text{cur}} \leftarrow E_n$
\STATE $\tilde{E}_n \leftarrow [E_n \;|\; \mathbf{1}_N]$ \COMMENT{Affine augmentation (Eq.~\ref{eq:augment})}
\FOR{$r = 1, \dots, R$}
    \item[] \hspace{1.5em}\textit{\textbf{OPCA} -- Confident Orthogonal Procrustes Alignment}
    \STATE $L \leftarrow E_{\text{cur}}\, T^\top$;\; $\hat{y} \leftarrow \arg\max_c L$;\; $\kappa \leftarrow \max_c L$;\; $\tau \leftarrow Q_{1-p}(\kappa)$
    \STATE $\mathcal{C} \leftarrow \{i : \kappa_i \geq \tau\}$;\; compute $M_n^c,\, \mu_n,\, \mu_t$ via Eq.~\ref{eq:centroid}
    \STATE $U, \Sigma, V^\top \leftarrow \mathrm{SVD}\bigl((M_n^{\mathcal{V}}{-}\mu_n)^\top (M_t^{\mathcal{V}}{-}\mu_t)\bigr)$;\; $R^* \leftarrow UV^\top$
    \STATE $s \leftarrow \mathrm{clamp}\bigl(\|M_t^{\mathcal{V}}{-}\mu_t\|_F \;/\; (\|M_n^{\mathcal{V}}{-}\mu_n\|_F{+}\epsilon),\, 0.8,\, 1.2\bigr)$
    \STATE $E' \leftarrow \mathrm{normalize}\bigl(s\,(E_{\text{cur}}{-}\mu_n)\,R^* + \mu_t\bigr)$ \COMMENT{Eq.~\ref{eq:opca}}
    \item[] \hspace{1.5em}\textit{\textbf{CCVD} -- Class-Conditioned Variance Deflation }
    \STATE Recompute $\hat{l},\, \kappa$ from $E'$;\; $\mathcal{F} \leftarrow \{i : \kappa_i \geq \mathrm{Pct}_{1-q}(\kappa)\}$
    \STATE Compute $S_W,\, S_B$ via Eqs.~\ref{eq:sw}--\ref{eq:sb}
    \STATE $V_K \leftarrow$ top-$K$ eigenvectors of $(S_B + \epsilon\, I)^{-1} S_W$ \COMMENT{Eq.~\ref{eq:fisher}}
    \STATE $E_{\text{ccvd}} \leftarrow \mathrm{normalize}\bigl(E'\,(I_d - V_K V_K^\top)\bigr)$ \COMMENT{Eq.~\ref{eq:ccvd}}
    \item[] \hspace{1.5em}\textit{\textbf{PCT} -- Per-Class Prototype Translation}
    \STATE $\hat{y} \leftarrow \arg\max_c (E_{\text{ccvd}}\, T^\top)$;\; compute $\mu_c^{\text{ccvd}}$ for each $c$
    \STATE $E_{\text{tgt}} \leftarrow \mathrm{normalize}\bigl(E_{\text{ccvd}} + \alpha\,(T[\hat{y}] - \mu^{\text{ccvd}}[\hat{y}])\bigr)$ \COMMENT{Eq.~\ref{eq:pct}}
    \item[] \hspace{1.5em}\textit{\textbf{ABR} -- Affine Bias Regression (compile to static map)}
    \STATE $W_{\text{aug}} \leftarrow (\tilde{E}_n^\top \tilde{E}_n + \lambda\, I_{d+1})^{-1}\, \tilde{E}_n^\top E_{\text{tgt}}$ \COMMENT{Eq.~\ref{eq:abr}}
    \STATE $E_{\text{cur}} \leftarrow \mathrm{normalize}(\tilde{E}_n\, W_{\text{aug}})$
\ENDFOR
\RETURN $E_{\text{cur}},\; W_{\text{aug}}$
\end{algorithmic}
\end{algorithm}

%%%%%%%%%%%%%%%%%%%%%%%%%%%%%%%%%%%%%%%%%%%%%
%%%%%%%%%%%%%%%%%%%%%%%%%%%%%%%%%%%%%%%%%%%%%%%%%%

Steps 1--3 produce a corrected target $E_{\text{tgt}}$ from the current round's embeddings. ABR distils all corrections into a single closed-form \emph{affine} map from the original noisy embeddings $E_n$, ensuring that the projection matrix is always a direct function of the observed data.

\paragraph{Affine Augmentation.}
To capture the translational component of the noise distortion alongside rotation and scaling, we augment the original noisy embeddings with a bias column:
\begin{equation}
\tilde{E}_n = \bigl[E_n \;\big|\; \mathbf{1}_N\bigr] \in \mathbb{R}^{N \times (d+1)},
\label{eq:augment}
\end{equation}
where $\mathbf{1}_N$ is a column vector of ones. This augmentation is computed once from the original $E_n$ and remains fixed across all calibration rounds.

\paragraph{Closed-Form Ridge Compilation.}
Let $W \in \mathbb{R}^{(d+1) \times d}$ denote the affine projection matrix to be estimated. $W$ minimises the regularised least-squares objective:
\begin{equation}
W_{\text{aug}}^{(r)} = \arg\min_{W \in \mathbb{R}^{(d+1) \times d}} \|\tilde{E}_n\, W - E_{\text{tgt}}\|_F^2 + \lambda \|W\|_F^2,
\label{eq:abr_obj}
\end{equation}
where $W$ is the optimisation variable over all $(d{+}1) \times d$ matrices and the subscript ``aug'' on the solution $W_{\text{aug}}^{(r)}$ emphasises that the map acts on the augmented (bias-appended) input. The unique closed-form solution is:
\begin{equation}
W_{\text{aug}}^{(r)} = \bigl(\tilde{E}_n^\top \tilde{E}_n + \lambda\, I_{d+1}\bigr)^{-1} \tilde{E}_n^\top E_{\text{tgt}}.
\label{eq:abr}
\end{equation}
The matrix $W_{\text{aug}} \in \mathbb{R}^{(d+1) \times d}$ decomposes row-wise as $W_{\text{aug}} = [W_{\text{lin}};\; b^\top]$: the upper $d \times d$ block $W_{\text{lin}}$ captures the linear component (rotation and scaling), while the final row $b \in \mathbb{R}^d$ captures the learned translational bias that directly absorbs the shift $\Delta_{\text{noise}}$ from Eq.~\ref{eq:anh}. This affine decomposition is what distinguishes \methodname\ from purely linear projection methods.

\begin{proposition}[Drift prevention via origin anchoring]
\label{prop:drift}
By regressing from the fixed $\tilde{E}_n$ at every round, ABR ensures that $W_{\text{aug}}^{(r)}$ maps directly from the original observation space. Consequently, successive corrections cannot amplify noise through repeated application of approximate inverse operations, as would occur if each round regressed from the previous round's output.
\end{proposition}

%%%%%%%%%%%%%%%%%%%%%%%%%%%%%%%%%%%%%%%%%%%%%%%%%%
\subsection{Iterative Calibration and Static Inferencing}
\label{sec:calibration}

\subsubsection*{Phase 1: Transductive Calibration (Batch-Dependent)}
The three geometric operations (OPCA, CCVD, Per-Class Translation) and the ABR compilation constitute a single correction pass. \methodname iterates this pass for $R{=}3$ rounds, bootstrapping progressively better pseudo-label estimates from the test batch alone. Each round's corrected $E_{\text{cur}}$ yields higher-quality pseudo-labels at the next round, which in turn produce more accurate centroid estimates (OPCA), a better-identified noise subspace (CCVD), and finer prototype translations. ABR anchors the ridge regression to the original $\tilde{E}_n$ at every round (Proposition~\ref{prop:drift}), preventing drift accumulation.

\paragraph{The Cross-Modal Anchor.}
In a unimodal TTA setting, the only self-supervised signal is internal batch consistency. \methodname instead exploits the pretrained audio-language model's text prototype matrix $T$, which encodes the semantic geometry of the class space in a noise-free modality. This provides an exogenous, label-free reference against which the noisy audio manifold is geometrically aligned at every round.

\subsubsection*{Phase 2: Batch-Independent Static Inference}
Once calibration is complete, $W_{\text{aug}}^{(R)}$ is frozen. Each new sample is adapted via a single $O(d^2)$ matrix-vector multiplication:
\begin{equation}
e_{\text{final}} = \mathrm{normalize}\bigl(\tilde{e}_{\text{noisy}} \cdot W_{\text{aug}}^{(R)}\bigr).
\label{eq:deploy}
\end{equation}
This requires zero gradients, zero batch statistics, and zero parameter updates, making \methodname suitable for edge inference on individual samples.
  %%%%%%%%%%%%%%%%%%%%%%%%%%%%%%%%%%%%%%%%%%%%%%%%%%%%%%%%%%%%%%%%%
\section{Experiments}
\label{sec:experiments}
We evaluate \methodname{} across multiple benchmark datasets and noise conditions to assess its robustness, generalization, and effectiveness under diverse distribution shifts. Beyond demonstrating strong performance under severe acoustic corruption, we also explicitly analyze the primary failure mode of the proposed framework, the Polyphonic Trap, and introduce Confidence-Aware Regression (CAR) as a targeted mitigation strategy, discussed in detail in \S~\ref{sec:polyphonic}.

%%%%%%%%%%%%%%%%%%%%%%%%%%%%%%%%%%%%%%%%%%%%%%%%%%%%%%%%%%%%%%%%%
% table 1
\begin{table}[t]
\caption{Comparison of \methodname{} against SoTA methods across multiple benchmarks. Best results are \textbf{bold}; second-best are \underline{underlined}. $^\dagger$ContextDA uses oracle noise-type prompts unavailable for ESC-50/DCASE. $^\ddagger$DCASE violates the Affine Noise Hypothesis via device mismatch rather than additive noise, evaluating robustness under assumption violation.}
\Description{Table comparing accuracy of PRISM and baseline methods across US8K, ESC-50, and DCASE datasets.}
\label{tab:main}
\centering
\small
\setlength{\tabcolsep}{6pt}
\renewcommand{\arraystretch}{0.92}
\resizebox{\linewidth}{!}{
\begin{tabular}{@{}lcccc@{}}
\toprule
 & \multicolumn{2}{c}{\textbf{US8K}} & \multicolumn{1}{c}{\textbf{ESC-50}} & \multicolumn{1}{c}{\textbf{DCASE$^\ddagger$}} \\
\cmidrule(lr){2-3} \cmidrule(lr){4-4} \cmidrule(lr){5-5}
\textbf{Method} & \textbf{Acc} & \textbf{F1} & \textbf{Acc} & \textbf{Acc} \\
\midrule
\multicolumn{5}{@{}l}{\textit{Zero-Shot Baseline}} \\
LAION-CLAP~\cite{wu2023laionclap} & 58.77 & 60.48 & 88.82 & 17.36 \\
\midrule
\multicolumn{5}{@{}l}{\textit{Fully Blind Transductive (Test-Time Adaptation)}} \\
TDA~\cite{karmanov2024efficient}  & 62.75             & 64.35             & 91.10 & \underline{17.38} \\
PCA++~\cite{wu2025pca}          & 67.88 & 68.57 & 77.57          & 15.10 \\
SubTTA~\cite{zeng2026subtta}      & 58.83             & 60.76             & 87.98          & 17.21 \\
\midrule
\multicolumn{5}{@{}l}{\textit{Oracle-Assisted (Noise-Guided Prompts)}} \\
ContextDA~\cite{acevedo2025domain}$^\dagger$ & 62.30 & ---  & ---   & --- \\
\midrule
\multicolumn{5}{@{}l}{\textit{Proposed Framework (Ours)}} \\
\textbf{\methodname + CAR}           & \underline{71.23}          & \underline{71.82}          & \underline{93.25}          & \textbf{17.70} \\
\textbf{\methodname}      & \textbf{71.71} & \textbf{72.36} & \textbf{93.39} & 15.63 \\

\bottomrule
\end{tabular}
}
\end{table}
%%%%%%%%%%%%%%%%%%%%%%%%%%%%%%%%%%%%%%%%%%%%%%%%%%%%%%%%%%%%%%%%%

%%%%%%%%%%%%%%%%%%%%%%%%%%%%%%%%%%%%%%%%%%%%%%%%%%%%%%%%%%%%%%%%%
\begin{table}[t]
\caption{Per-environment accuracy (\%) on US8K (10-fold CV).}
\Description{Table showing accuracy breakdown for 10 individual noise environments on US8K.}
\label{tab:perbg}
\centering
\small
\setlength{\tabcolsep}{6pt}
\renewcommand{\arraystretch}{0.92}
\resizebox{\linewidth}{!}{
\begin{tabular}{@{}lcccc|cc@{}}
\toprule
\textbf{Background} & \textbf{Zero-Shot} & \textbf{SubTTA} & \textbf{PCA++} & \textbf{TDA} & \textbf{\methodname~+~CAR} & \textbf{\methodname} \\
\midrule
shopping\_mall   & 49.43 & 49.19 & 64.67 & 55.19 & 68.47 & \textbf{68.53} \\
street\_ped.     & 53.05 & 52.95 & 66.81 & 58.37 & 71.17 & \textbf{71.50} \\
public\_square   & 54.11 & 54.25 & 66.56 & 58.54 & 70.66 & \textbf{70.85} \\
airport          & 56.02 & 56.16 & 67.02 & 60.89 & \textbf{70.76} & 70.67 \\
metro\_station   & 58.38 & 58.49 & 69.38 & 62.57 & 71.15 & \textbf{71.34} \\
street\_traffic  & 60.54 & 60.97 & 67.25 & 64.19 & 71.76 & \textbf{71.87} \\
park             & 63.83 & 64.11 & 69.22 & 67.19 & \textbf{73.85} & 73.69 \\
bus              & 64.22 & 64.36 & 67.53 & 66.87 & 73.35 & \textbf{73.60} \\
tram             & 63.99 & 63.93 & 68.69 & 66.67 & \textbf{72.70} & 72.65 \\
metro            & 64.10 & 63.91 & 71.69 & 67.01 & 72.41 & \textbf{72.42} \\
\midrule
\textbf{Average} & 58.77 & 58.83 & 67.88 & 62.75 & 71.23 & \textbf{71.71} \\
\bottomrule
\end{tabular}
}
\end{table}
%%%%%%%%%%%%%%%%%%%%%%%%%%%%%%%%%%%%%%%%%%%%%%%%%%%%%%%%%%%%%%%%%

%%%%%%%%%%%%%%%%%%%%%%%%%%%%%%%%%%%%%%%%%%%%%%%%%%
\subsection{Experimental Setup}
\label{sec:setup}

%%%%%%%%%%%%%%%%%%%%%%%%%%%%%%%%%%%%%%%%%%%%%%%%%%%%%%%%%%%%%%%%%
% table 3
\begin{table}[t]
\caption{Per-environment accuracy (\%) on ESC-50 (5-fold CV).}
\Description{Table showing accuracy breakdown for 10 individual noise environments on ESC-50.}
\label{tab:esc50_perbg}
\centering
\small
\setlength{\tabcolsep}{6pt}
\renewcommand{\arraystretch}{0.92}
\resizebox{\linewidth}{!}{
\begin{tabular}{@{}lcccc|cc@{}}
\toprule
\textbf{Background} & \textbf{Zero-Shot} & \textbf{SubTTA} & \textbf{PCA++} & \textbf{TDA} & \textbf{\methodname~+~CAR} & \textbf{\methodname} \\
\midrule
shopping\_mall   & 86.45 & 86.20 & 74.40 & 88.95 & 91.40 & \textbf{91.60} \\
street\_ped.     & 86.70 & 86.90 & 74.20 & 88.90 & \textbf{92.65} & 92.60 \\
public\_square   & 87.70 & 88.05 & 78.70 & 90.40 & 92.60 & \textbf{92.70} \\
airport          & 86.35 & 86.05 & 75.25 & 88.90 & 91.55 & \textbf{91.60} \\
metro\_station   & 89.30 & 89.10 & 80.50 & 91.30 & 92.95 & \textbf{93.05} \\
street\_traffic  & 89.00 & 89.20 & 81.50 & 91.70 & 93.70 & \textbf{94.10} \\
park             & 90.15 & 90.45 & 74.35 & 92.65 & 94.80 & \textbf{95.05} \\
bus              & 91.90 & 92.40 & 76.75 & 93.55 & 94.90 & \textbf{95.10} \\
tram             & 90.95 & 91.30 & 80.25 & 92.70 & 94.35 & \textbf{94.40} \\
metro            & 89.75 & 90.25 & 79.75 & 91.95 & 93.55 & \textbf{93.70} \\
\midrule
\textbf{Average} & 88.82 & 88.99 & 77.57 & 91.10 & 93.25 & \textbf{93.39} \\
\bottomrule
\end{tabular}
}
\end{table}
%%%%%%%%%%%%%%%%%%%%%%%%%%%%%%%%%%%%%%%%%%%%%%%%%%%%%%%%%%%%%%%%%

%%%%%%%%%%%%%%%%%%%%%%%%%%%%%%%%%%%%%%%%%%%%%%%%%%%%%%%%%%%%%%%%%
%%%% table 4
\begin{table}[t]
\caption{Accuracy (\%) on US8K across varying SNR values. Red-shaded rows denote low-SNR regimes where noise dominates foreground events; $\dag$ marks moderate noise levels, and $\star$ indicates anomalous performance drops.}
\Description{Table showing accuracy across different SNR levels from -6 dB to clean on US8K.}
\label{tab:snr}
\centering
\small
\setlength{\tabcolsep}{6pt}
\renewcommand{\arraystretch}{0.92}
\resizebox{\linewidth}{!}{
\begin{tabular}{@{}lcccc|cc@{}}
\toprule
\textbf{SNR} & \textbf{Zero-Shot} & \textbf{SubTTA} & \textbf{PCA++} & \textbf{TDA} & \textbf{\methodname~+~CAR} & \textbf{\methodname} \\
\midrule
\rowcolor{red!8} $-$6\,dB & 32.69 & 33.02 & 37.55 & 37.47 & 54.03 & \textbf{57.45} \\
\rowcolor{red!5} $-$3\,dB & 49.70 & 49.99 & 56.56 & 54.32 & 64.76 & \textbf{66.49} \\
\rowcolor{red!3} $\phantom{-}$0\,dB & 64.85 & 64.77 & 70.62 & 68.28 & 74.53 & \textbf{74.68} \\
\midrule
$\phantom{-}$3\,dB  & 72.38 & 72.33 & 76.76 & 75.45 & 80.29 & \textbf{81.24} \\
$\phantom{-}$6\,dB$^\dag$  & 75.54 & 75.43 & 77.80 & 78.34 & 82.71 & \textbf{83.79} \\
$\phantom{-}$8\,dB$^\dag$  & 76.59 & 76.53 & 77.58 & 79.17 & 83.65 & \textbf{84.78} \\
$\phantom{}$10\,dB$^\dag$ & 77.12 & 77.18 & 76.14$^\star$ & 79.73 & 84.40 & \textbf{84.84} \\
$\phantom{}$15\,dB  & 77.93 & 77.98 & 79.02 & 80.27 & 85.13 & \textbf{85.35} \\
Clean   & 78.39 & 78.40 & 77.10$^\star$ & 80.40 & 85.47 & \textbf{85.57} \\
\midrule
\rowcolor{gray!25} \multicolumn{7}{@{}l}{\textit{Real Conditions (From Table~\ref{tab:main})}} \\
\rowcolor{gray!25} Mixed SNR (Avg) & 58.77 & 58.83 & 67.88 & 62.75 & 71.23 & \textbf{71.71}\\
\bottomrule
\end{tabular}
}
\end{table}
%%%%%%%%%%%%%%%%%%%%%%%%%%%%%%%%%%%%%%%%%%%%%%%%%%%%%%%%%%%%%%%%%

\subsubsection{Datasets}
\label{sec:datasets}
We evaluate on three benchmark datasets: \textbf{(a) UrbanSound8K (US8K)}~\cite{salamon2014urbansound8k}, containing 8{,}732 clips across 10 urban sound classes under 10-fold cross-validation, where we inject background noise from 10 environments in the TAU Urban Acoustic Scenes 2019 corpus~\cite{mesaros2018tau} at varying intensity levels, yielding 87{,}320 evaluation instances and forming the most comprehensive transductive evaluation regime in the literature; \textbf{(b) ESC-50}~\cite{piczak2015esc50}, containing 2{,}000 clips across 50 environmental sound classes under 5-fold cross-validation, with noise injected from the same 10 TAU backgrounds, matching the US8K evaluation protocol; and \textbf{(c) DCASE/TAU 2019}~\cite{mesaros2018tau}, containing 14{,}400 clips across 10 scene categories, where domain shift arises naturally from real-world recording device mismatch rather than synthetic noise injection, providing a complementary out-of-distribution evaluation setting.

\subsubsection{Implementation Details}
\label{sec:impl}
All experiments use LAION-CLAP (checkpoint \texttt{630k-audioset-fusion-best.pt})~\cite{wu2023laionclap}, run on a workstation with an Intel Xeon Gold 5320 CPU and a single NVIDIA A100 80\,GB GPU. Text prototypes are computed by averaging embeddings from 20 diverse prompt templates per class (e.g., \textit{``This is a sound of \{\}''}, \textit{``An audio clip of \{\}''}) and $\ell_2$-normalising. \methodname{} performs three iterative rounds of OPCA, CCVD, and ABR using $K{=}60$, confidence retention thresholds $p{=}0.8$ and $q{=}0.7$, ridge coefficient $\lambda{=}0.01$, and prototype translation strength $\alpha{=}0.3$. The CAR variant additionally employs confidence-aware regression ($\gamma{=}10$) within the transductive calibration loop to mitigate polyphonic failure modes (Section~\ref{sec:polyphonic}). All hyperparameters are fixed across every fold, environment, and dataset without per-condition tuning.

\noindent\textbf{Calibration Protocol.}
Following standard transductive TTA evaluation~\cite{karmanov2024efficient, wu2025pca}, $W_{\text{aug}}$ is estimated from all unlabeled noisy embeddings available in each cross-validation split (both training and test partitions combined). No ground-truth labels are used during calibration; only the unlabeled embeddings and frozen text prototypes $T$. The same calibrated samples are subsequently evaluated, consistent with the transductive protocol used by all baselines. Calibration is performed independently per fold and per background environment. For fair comparison, all baselines are reproduced and evaluated under an identical protocol.

\subsubsection{Baselines}
\label{sec:baselines}

We compare \methodname{} against both \textit{Fully Blind Transductive} and \textit{Oracle-Assisted Prompt Guided} adaptation methods under a unified protocol using identical $\ell_2$-normalized LAION-CLAP embeddings, shared prompt-ensemble text prototypes $\mathbf{T}$, and identical fold structures without access to ground-truth labels. Evaluated blind baselines include: \textit{Zero-Shot (LAION-CLAP)}, which performs direct cosine-similarity classification without adaptation; \textit{SubTTA}~\cite{zeng2026subtta}, which aligns acoustic and text subspaces via chordal-distance minimization; \textit{PCA++}~\cite{wu2025pca}, which performs transductive subspace alignment through the generalized eigenproblem $\mathbf{S}_W^{-1}\mathbf{S}_B$ ($K{=}60$, 3 rounds); and \textit{TDA}~\cite{karmanov2024efficient}, a confidence-weighted key-value cache adaptation method ($\alpha{=}0.5$, 3 rounds). Since SubTTA, PCA++, and TDA were originally designed for CLIP-based vision-language models, we adapt them directly to CLAP embeddings due to the shared dual-encoder contrastive architecture. We additionally evaluate the oracle-assisted \textit{ContextDA}$^\dagger$~\cite{acevedo2025domain}, which receives the background noise type as an inference-time oracle prompt and thus serves as an upper bound for prompt-guided adaptation. For fair comparison, we reproduce ContextDA under our evaluation protocol (10 backgrounds, mixed SNR, 10-fold CV), whereas the original work evaluated only 6 backgrounds at fixed 6--10\,dB SNR; direct per-SNR comparison is provided in Table~\ref{tab:snr}.

%%%%%%%%%%%%%%%%%%%%%%%%%%%%%%%%%%%%%%%%%%%%%%%%%%%%%%%%%%%%%%%%%
\subsection{Results and Discussion}\label{sec:results}

%%%%%%%%%%%%%%%%%%%%%%%%%%%%%%%%%%%%%%%%%%%%%%%%%%%%%%%%%%%%%%%%%
\begin{figure*}[t]
\centering
\includegraphics[width=0.75\linewidth]{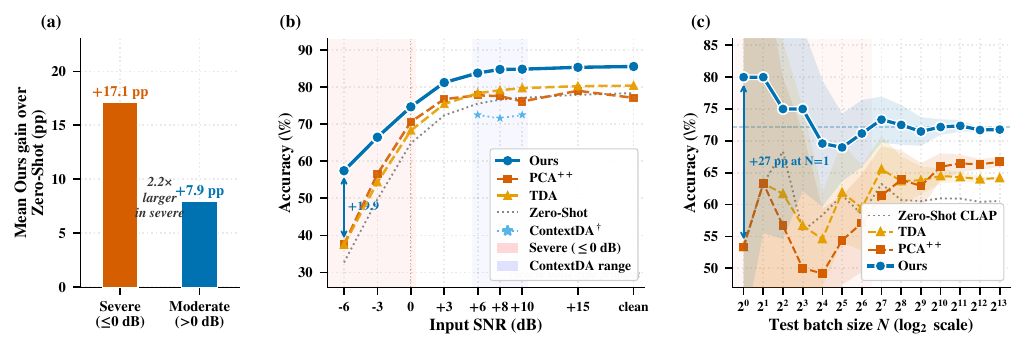}
\caption{\textbf{Operational Robustness of \methodname.}
  \textbf{(a)}~Mean gain over zero-shot in the severe regime and in moderate conditions.
  \textbf{(b)}~Accuracy vs.\ input SNR on US8K.
  \textbf{(c)}~Accuracy vs.\ test batch size $N$. \methodname\ calibrates its projection matrix once offline.}
\label{fig:robustness}
\Description{Three side-by-side plots. Left: average accuracy gains for severe vs. moderate noise conditions. Middle: accuracy vs. SNR from -6 dB to clean. Right: accuracy vs. test batch size, showing PRISM holds up even for small batches.}
\end{figure*}
%%%%%%%%%%%%%%%%%%%%%%%%%%%%%%%%%%%%%%%%%%%%%%%%%%%%%%%%%%%%%%%%%

%%%%%%%%%%%%%%%%%%%%%%%%%%%%%%%%%%%%%%%%%%%%%%%%%%%%%%%%%%%%%%%%%
\begin{figure*}[t]
\centering
    \includegraphics[width=\linewidth]{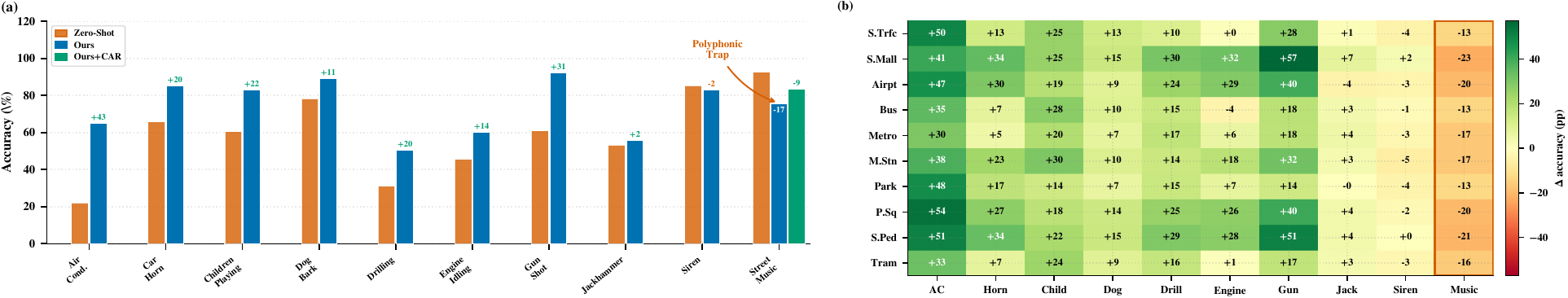}
\caption{\textbf{Diagnostic Analysis of \methodname.}
\textbf{Left:} Per-class accuracy vs.\ zero-shot CLAP. Spectrally sparse classes gain substantially, confirming CCVD separates noise from signal for impulsive classes.
\textbf{Right:} Cross-environment heatmap ($\Delta$ Acc = \methodname\ $-$ ZS). Consistent improvement across 8 classes validates the Affine Noise Hypothesis.}
\label{fig:diagnostic}
\Description{Two panels. Left: a bar chart of per-class accuracy showing most classes improve under PRISM, but street music drops. Right: a heatmap of per-class accuracy change across all 10 noise environments, confirming that street music degrades uniformly regardless of background.}
\end{figure*}
%%%%%%%%%%%%%%%%%%%%%%%%%%%%%%%%%%%%%%%%%%%%%%%%%%%%%%%%%%%%%%%%%

%%%%%%%%%%%%%%%%%%%%%%%%%%%%%%%%%%%%%%%%%%%%%%
\subsubsection{Quantitative Results}
\label{sec:quantitative_results}
\methodname{} establishes a new SoTA on every additive-noise benchmark in Table~\ref{tab:main}. On US8K, \methodname\ achieves \textbf{71.71\%} accuracy and \textbf{72.36\%} Macro-F1 under the most comprehensive evaluation in the literature (10 backgrounds $\times$ 10-fold CV), a $+$12.94 percentage point (pp) absolute gain over the zero-shot baseline.
Against blind competitors, \methodname\ outperforms the strongest
training-free method (PCA++) by $+$3.83\,pp and the cache-based TDA by
$+$8.96\,pp; all evaluated under the exact same transductive protocol.
Notably, zero-shot CLAP and standard entropy minimization yield nearly
identical accuracy ($\sim$58.8\%), confirming that na\"ive entropy
minimization provides essentially zero benefit under severe acoustic noise,
validating our geometric approach.
Furthermore, \methodname\ surpasses the oracle-assisted ContextDA by
$\mathbf{+9.41}$\,pp, even though ContextDA receives privileged noise-type
prompts unavailable in real inference scenarios.
On ESC-50, \methodname\ achieves \textbf{93.39\%}, a $+$2.29\,pp gain over
TDA (91.10\%) and notably a $+$15.82\,pp improvement over PCA++
(77.57\%), which \emph{regresses} below zero-shot on this benchmark.
This catastrophic failure of PCA++ on ESC-50 is not a coincidence; it is a
direct consequence of the Spectral Over-Alignment Dilemma analyzed in
Section~\ref{sec:dilemma}.

Table~\ref{tab:perbg} and Table~\ref{tab:esc50_perbg} provide the per-environment breakdown on US8K
and ESC-50, respectively, confirming that \methodname's superiority is consistent across
all 10 noise backgrounds on both benchmarks, with ESC-50 gains ranging from
$+$3.20\,pp (bus) to $+$5.90\,pp (street\_pedestrian) over zero-shot.
Critically, PCA++ \emph{degrades} accuracy by 7--16\,pp
across every ESC-50 environment ($-$11.25\,pp average), confirming the
Spectral Over-Alignment Dilemma is environment- and dataset-agnostic.

\noindent\textbf{Environment Robustness.}
Table~\ref{tab:perbg} demonstrates that \methodname's superiority is not
driven by a few easy environments: it achieves the highest accuracy in
\emph{every single background}, with per-environment gains ranging from
$+$8.32\,pp (metro) to $+$19.10\,pp (shopping mall) over zero-shot.
Notably, the hardest environment for zero-shot (shopping\_mall: 49.43\%)
sees the largest absolute gain, confirming that CCVD's subspace projection
is most effective precisely where noise dominance is strongest.

\noindent\textbf{Severe Acoustic Regime ($\leq\!0$\,dB SNR).}
While prior work, including the closest oracle-assisted baseline
ContextDA~\cite{acevedo2025domain}, restricts evaluation to 6--10\,dB SNR,
real-world inference such as industrial monitoring, crowd sensing, and
mobile audio search frequently encounter conditions where the noise floor
\emph{dominates} the target signal ($\leq\!0$\,dB).
We extend the SNR sweep into this \emph{severe} regime for the first time.
At $-$6\,dB, zero-shot CLAP collapses to 32.69\% and PCA++ reaches only
37.55\%, while \methodname\ achieves \textbf{57.45\%}, a $+$24.76\,pp gain
over zero-shot and $+$19.90\,pp over PCA++, averaged over 4 backgrounds and
10 cross-validation folds.
In particular, ContextDA's best oracle-assisted accuracy at 6\,dB ($\sim$72.5\%)
is \emph{already 11.3\,pp below} \methodname's 83.79\% at the same SNR,
achieved entirely without noise-type annotation.
Panel (a) of Figure~\ref{fig:robustness} confirms that the \methodname\
advantage over zero-shot is monotonically largest in the severe regime,
providing the strongest empirical validation of the Affine Noise Hypothesis:
the more dominant the additive acoustic distortion, the more low-rank its
embedding-space signature, and the more effectively CCVD isolates and
removes it.

\noindent\textbf{DCASE: Robustness under Assumption Violation.}
The DCASE/TAU benchmark involves device mismatch, an organic domain shift that does \emph{not} produce the low-rank additive distortion predicted by the Affine Noise Hypothesis. As expected, all methods struggle: Zero-Shot (17.36\%), TDA (17.38\%), and SubTTA (17.21\%). Because the low-rank assumption is violated, hard geometric projections actively destroy signal, as seen by the degradation in PCA++ (15.10\%) and base \methodname\ (15.63\%). Critically, \methodname~+~CAR detects this instability and acts as a geometric safety parachute. It is the \emph{only} method that achieves a meaningful improvement over the zero-shot baseline (17.70\%), confirming that confidence-aware regression successfully prevents the catastrophic failure modes of rigid transductive alignment.

%%%%%%%%%%%%%%%%%%%%%%%%%%%%%%%%%%%%%%%%%%%%%%
\subsubsection{Qualitative Results}
\label{sec:qualitative_results}

% %%%%%%%%%%%%%%%%%%%%%%%%%%%%%%%%%%%%%%%%%%%%%%%%%%
% \begin{figure*}[t]
% \centering
%     \includegraphics[width=\linewidth]{figures/fig_ablation_sensitivity_combined.pdf}
% \caption{Sensitivity Analysis on US8K.
%   \textbf{Left (a,b):} Overall accuracy per method and incremental component ablation.
%   \textbf{Right (c--e):} Hyperparameter sensitivity across Noise Subspace rank $K \in [30, 100]$, Shift Strength $\alpha \in [0, 1]$, and Ridge Penalty $\lambda \in [10^{-3}, 10^{-1}]$.}
% \label{fig:ablation_sensitivity}
% \Description{Five panels. Top row: two bar charts comparing methods and PRISM's incremental components. Bottom row: three line plots showing how accuracy changes with noise-subspace rank K, prototype translation strength, and ridge penalty.}
% \end{figure*}
% %%%%%%%%%%%%%%%%%%%%%%%%%%%%%%%%%%%%%%%%%%%%%%%%%%

%%%%%%%%%%%%%%%%%%%%%%%%%%%%%%%%%%%%%%%%%%%%%%%%%%
\begin{figure*}[t]
\centering
\begin{minipage}[t]{0.405\textwidth}
    \centering
    \includegraphics[width=\linewidth, height=3.2cm, trim=0 0 0 2.5mm, clip ]{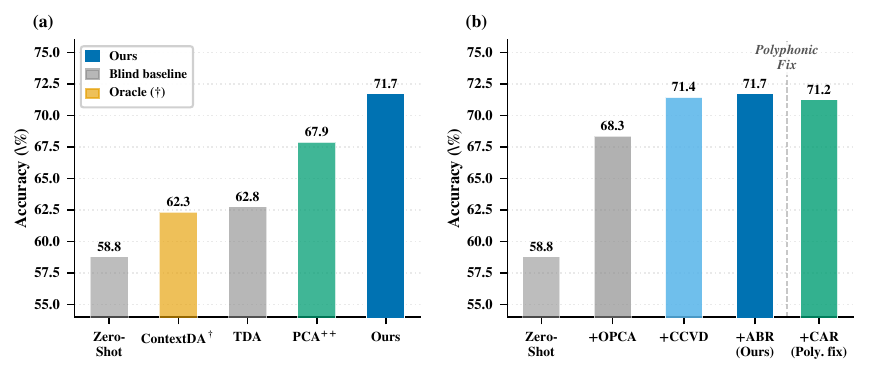}
\end{minipage}
% \hfill
\begin{minipage}[t]{0.585\textwidth}
    \centering
    \includegraphics[width=\linewidth, height=3.2cm]{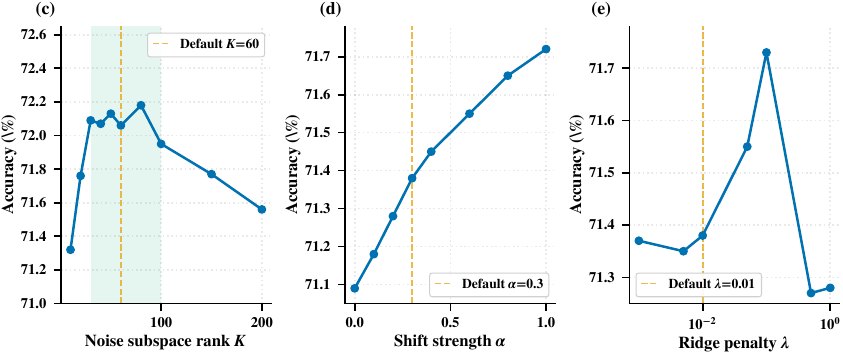}
\end{minipage}
\caption{Sensitivity Analysis on US8K.
  \textbf{Left (a,b):} Overall accuracy per method and incremental component ablation. 
  \textbf{Right (c--e):} Hyperparameter sensitivity across Noise Subspace rank $K \in [30, 100]$, Shift Strength $\alpha \in [0, 1]$, and Ridge Penalty $\lambda \in [10^{-3}, 10^{-1}]$.}
\label{fig:ablation_sensitivity}
\vspace{-2mm}
\end{figure*}
%%%%%%%%%%%%%%%%%%%%%%%%%%%%%%%%%%%%%%%%%%%%%%%%%%

%%%%%%%%%%%%%%%%%%%%%%%%%%%%%%%%%%%%%%%%%%%%%%%%%%
\begin{table}[t]
\centering
\caption{Incremental component ablation on US8K.
Each row adds one component to the previous.}
\Description{Table showing accuracy and F1 score for incremental additions of OPCA, CCVD, and ABR.}
\label{tab:ablation}
\scriptsize
\renewcommand{\arraystretch}{0.92}
\resizebox{0.95\linewidth}{!}{
\begin{tabular}{l l r r}
\toprule
\textbf{Config} & \textbf{Component Added} & \textbf{Acc\,(\%)} & \textbf{F1\,(\%)} \\
\midrule
Zero-Shot            & ---                     & 58.77 & 60.48 \\
+OPCA                & Cross-modal alignment   & 68.35 & 68.87 \\
+CCVD                & Noise subspace removal  & 71.44 & 72.02 \\
\textbf{\methodname} & Affine bias correction  & \textbf{71.71} & \textbf{72.36} \\
\midrule
\methodname~+~CAR      & Polyphonic mitigation   & 71.23 & 71.82 \\
\bottomrule
\end{tabular}
}
\end{table}
%%%%%%%%%%%%%%%%%%%%%%%%%%%%%%%%%%%%%%%%%%%%%%%%%%

%%%%%%%%%%%%%%%%%%%%%%%%%%%%%%%%%%%%%%%%%%%%%%%%%%
\begin{table}[t]
\centering
\caption{Computational cost comparison across SoTA.}
\Description{Table comparing precomputation time, inference time, and parameter counts across methods.}
\label{tab:compute}
\small
\renewcommand{\arraystretch}{0.92}
\resizebox{0.95\linewidth}{!}{
\begin{tabular}{l r r c l}
\toprule
\textbf{Method} & \textbf{Precomp.} & \textbf{Inference} & \textbf{Grad.} & \textbf{Params} \\
\midrule
CoNMix              & N/A              & $\sim$50\,ms    & \cmark & Millions \\
TDA                 & $\sim$5\,ms/samp & $\sim$2\,ms     & \xmark & Cache \\
PCA++               & $\sim$80\,ms     & $\sim$0.008\,ms & \xmark & None \\
\textbf{\methodname} & \textbf{311\,ms} & \textbf{0.0009\,ms} & \xmark & \textbf{None} \\
\bottomrule
\end{tabular}
}
\end{table}
%%%%%%%%%%%%%%%%%%%%%%%%%%%%%%%%%%%%%%%%%%%%%%%%%%

In this section, we analyze the qualitative behavior of \methodname{} through diagnostic visualizations of robustness, class-wise performance, and geometric failure modes under severe acoustic shift. These analyses provide additional insight into the operational characteristics of the proposed framework beyond aggregate quantitative metrics.

\noindent\textbf{Operational Desiderata.} 
Figure~\ref{fig:teaser_master}b demonstrates that \methodname\ is the \emph{only}
method that is training-free, closed-form, blind to target labels and noise-type annotations, and batch-independent during inference.

\subsubsection{Spectral Over-Alignment Dilemma}
\label{sec:dilemma}
Figure~\ref{fig:robustness}(b) reveals a failure mode of hard uniformity-based
alignment: at 10\,dB SNR, PCA++ (76.14\%) drops \emph{below} the zero-shot
baseline (77.12\%), and this inversion recurs at clean audio
(77.10\% vs.\ 78.39\%), revealing a non-monotonic instability.
We term this the \textit{Spectral Over-Alignment Dilemma}.

\noindent\textbf{Mechanism.} 
PCA++ identifies noise directions by solving the LDA eigenproblem
$\mathbf{S}_W^{-1}\mathbf{S}_B$ and then projects features onto the
complement of those directions.
At high noise levels ($\leq$6\,dB), this beneficially discards the dominant
noise variance directions.
However, at moderate SNR ($\geq$10\,dB), the audio signal retains rich
intra-class geometric structure (spectral harmonics, timbral variation,
and temporal modulation patterns), all of which contribute discriminative
variance.
Because PCA++ applies the same complement projection regardless of SNR, it
indiscriminately erases these \emph{signal-bearing} directions, collapsing the
embedding geometry needed for classification.

\noindent\textbf{Why \methodname\ is immune?}
CCVD solves the \emph{inverse} Fisher ratio $F = (\mathbf{S}_B + \epsilon I)^{-1}\mathbf{S}_W$,
whose leading eigenvectors maximise within-class spread relative to
between-class discriminability which is the geometric signature of noise, not signal.
The complement projector $I_d - V_K V_K^\top$ then discards exactly
these noise directions while leaving all signal-bearing structure intact.
Table~\ref{tab:snr} confirms \methodname\ achieves the highest accuracy at
every SNR level with strictly monotonically increasing performance from $-$6\,dB to
clean audio.

%%%%%%%%%%%%%%%%%%%%%%%%%%%%%%%%%%%%%%%%%%%%%%%%%%%%%%%%%%%%%%%%%
\subsection{Ablation Study}
\label{sec:ablation}

%%%%%%%%%%%%%%%%%%%%%%%%%%%%%%%%%%%%%%%%%%%%%%%%%%%%%%%%
\begin{table}[t]
\caption{Per-class accuracy (\%) on US8K averaged across all 10 backgrounds.
  \methodname~+~CAR \emph{partially mitigates} the Polyphonic Trap for street music
  via confidence-aware regression.}
\Description{Table showing per-class accuracy across 10 classes on US8K for Zero-Shot, PRISM+CAR, and PRISM.}
\label{tab:perclass}
\centering
\small
\setlength{\tabcolsep}{12pt}
\renewcommand{\arraystretch}{0.92}
\begin{tabular}{@{}lccc@{}}
\toprule
\textbf{Class} & \textbf{Zero-Shot} & \textbf{\methodname~+~CAR} & \textbf{\methodname} \\
\midrule
air\_conditioner  & 22.06 & 58.35 & \textbf{64.77} \\
gun\_shot         & 60.80 & 86.36 & \textbf{92.14} \\
children\_playing & 60.50 & 80.45 & \textbf{82.80} \\
car\_horn         & 65.62 & 82.63 & \textbf{85.36} \\
drilling          & 30.97 & 49.42 & \textbf{50.60} \\
engine\_idling    & 45.55 & \textbf{62.01} & 59.95 \\
dog\_bark         & 78.20 & 85.87 & \textbf{88.93} \\
jackhammer        & 53.15 & \textbf{56.16} & 55.65 \\
siren             & \textbf{85.37} & 84.48 & 82.98 \\
street\_music     & \textbf{92.53} & \underline{83.49} & 75.33 \\
\midrule
\textbf{Overall}  & 58.77 & 71.23 & \textbf{71.71} \\
\bottomrule
\end{tabular}
\end{table}
%%%%%%%%%%%%%%%%%%%%%%%%%%%%%%%%%%%%%%%%%%%%%%%%%%%%%%%%

\noindent\textbf{Component Ablation.}
Table~\ref{tab:ablation} and Figure~\ref{fig:ablation_sensitivity} isolate the
contribution of each \methodname\ component in a strict incremental protocol
(all evaluated on the full US8K, 10 backgrounds, 10-fold CV).

\noindent\textit{OPCA} ($+$9.58\,pp): Orthogonal Procrustes cross-modal
alignment is the dominant transformation.
It establishes the correct geometric correspondence between audio and text
prototype manifolds, correcting the systematic bias introduced by background
noise without any class label information.

\noindent\textit{CCVD} ($+$3.09\,pp): Class-Conditioned Variance Deflation
delivers the second-largest gain by removing within-class noise variance in
a class-conditioned subspace.
This confirms the core claim of the Affine Noise Hypothesis: background noise
acts as an additive affine shift whose geometric signature is separable from
the signal subspace.

\noindent\textit{ABR} ($+$0.27\,pp): Affine Bias Regression provides a
final closed-form correction via ridge regression, absorbing any residual
covariate shift not captured by the preceding subspace operations.

\noindent\textbf{Hyperparameter Sensitivity.}
Figure~\ref{fig:ablation_sensitivity} demonstrates that \methodname\ is not brittle to
hyperparameter choices.
The noise subspace rank $K$ exhibits a wide robust plateau: accuracy varies
by $<$0.1\,pp across $K \in [30, 100]$, confirming that the noise subspace
is geometrically well-defined.
Shift strength $\alpha$ controls how aggressively the affine correction is
applied; while accuracy improves monotonically with $\alpha$, total
variation is $<$0.63\,pp across the full $[0,1]$ range.  We adopt
$\alpha{=}0.3$ as a conservative default (71.38\%), which sits only
0.34\,pp below the $\alpha{=}1.0$ peak, a near-insensitive choice that
remains robust when the noise distribution at test time differs from the
calibration set.
The ridge penalty $\lambda$ peaks at $10^{-1}$ (71.73\%) and remains stable
across two orders of magnitude; we use $\lambda{=}0.01$ as our conservative default
(71.38\%), which sits only 0.35\,pp below the peak and provides better
generalization to unseen noise distributions.

\noindent\textbf{Computational Efficiency \& Batch Independence.}
After a one-time precomputation of 311\,ms, each test sample is adapted via
a single closed-form matrix-vector multiply in \textbf{0.0009\,ms}, which is $9\times$
faster than PCA++ and orders of magnitude faster than any gradient-based
method (Table~\ref{tab:compute}; Figure~\ref{fig:teaser_master}).
Critically, \methodname\ achieves its $+$12.94\,pp accuracy gain over zero-shot
\emph{without any speed penalty}: the inference-time latency of 0.0009\,ms is
faster than even the un-adapted zero-shot baseline.
In contrast, TTA methods sacrifice 3--5 orders of magnitude in speed
(CoNMix: 50\,ms; TDA: 2\,ms) for marginal or even
negative accuracy gains over zero-shot.
Crucially, Figure~\ref{fig:robustness}(c) confirms \emph{true} inference batch independence:
while transductive baselines collapse below 65\% for $N \le 64$,
\methodname\ maintains $72.2 \pm 0.6$\% across all batch sizes and achieves
\textbf{80.0\%} at $N{=}1$, since the offline-calibrated projection matrix
requires only a short warm-up buffer ($N{=}128$--$512$ samples) and applies
to live audio streams one sample at a time.

%%%%%%%%%%%%%%%%%%%%%%%%%%%%%%%%%%%%%%%%%%%%%%%%%%%%%%%%%%%%%%%%%%%%%%%%%%%%%%%%
\section{Limitations}
\label{sec:polyphonic}

In this section, we analyze a principled limitation of \methodname{} that emerges for highly polyphonic sound classes despite its state-of-the-art overall performance. We further characterize the geometric origin of this failure mode and introduce a mathematically grounded mitigation strategy.

%%%%%%%%%%%%%%%%%%%%%%%%%%%%%%%%%%%%%%%%%%%%%%%%%%%%%%%%
\subsection{Problem: When Semantic Variance Resembles Noise}

The denoising assumptions underlying CCVD remain highly effective for spectrally sparse and impulsive events, but become unstable for acoustically dense polyphonic classes whose intrinsic semantic structure exhibits high spectral variance. The top panel of Figure~\ref{fig:diagnostic} reveals a sharp dichotomy: impulse-like, spectrally sparse classes gain substantially under CCVD (air\_conditioner: $+$42.71\,pp; gun\_shot: $+$31.34\,pp; children\_playing: $+$22.30\,pp), while \emph{street\_music} suffers a $-$17.20\,pp collapse from 92.53\% to 75.33\%.

We identify this as the \textbf{Polyphonic Trap}:
\textit{
``Polyphonic sound classes encode semantic identity through a
superposition of harmonics spanning broad spectral bands.
In the CLAP embedding space, these harmonics generate high within-class
variance that overlaps geometrically with the low-frequency noise directions
targeted by CCVD. The class-conditioned variance deflation therefore
inadvertently erases signal-bearing spectral components, collapsing the
embedding representation of the class.''
}

Critically, Figure~\ref{fig:diagnostic} confirms that the Polyphonic Trap is not an acoustic artifact but a geometric failure mode: \emph{street\_music} degrades uniformly across all 10 environments ($-$13 to $-$23\,pp), while most other classes consistently improve. Such environment-independent regularity is the empirical signature of subspace overlap in the CLAP latent space, as predicted by the Rank Overlap Constraint ($\mathrm{rank}(\Sigma_{\mathrm{class}}) + K > d$). Table~\ref{tab:perclass} further shows that zero-shot CLAP already achieves 92.53\% on \emph{street\_music}, confirming that CCVD disrupts rich polyphonic structure, whereas the siren class exhibits only a mild overlap effect ($-$2.39\,pp), recovered to 84.48\% by CAR.

%%%%%%%%%%%%%%%%%%%%%%%%%%%%%%%%%%%%%%%%%%%%%%%%%%%%%%%%
\subsection{Solution: Confidence-Aware Regression}

To mitigate the Polyphonic Trap, we introduce \textit{Confidence-Aware Regression (CAR)}, a confidence-gated interpolation designed to preserve semantically informative variance for acoustically dense polyphonic classes. For each sample embedding $\mathbf{e}_i$ predicted as class $\hat{c}$, let $\bar{\sigma}_{\hat{c}}$ denote the mean max-logit confidence of all samples assigned to class $\hat{c}$, and let $m$ denote the median confidence across all predicted classes. CAR computes a confidence-based weight and the resulting interpolation target:
\begin{equation}
\begin{aligned}
  w_i &= \sigma\!\bigl(\gamma\,(m - \bar{\sigma}_{\hat{c}})\bigr),\\
  \tilde{\mathbf{e}}_i
      &= w_i\,\mathbf{e}_i^{\mathrm{noisy}}
       + (1-w_i)\,\mathbf{e}_i^{\mathrm{PRISM}}.
\end{aligned}
  \label{eq:car}
\end{equation}
where $\sigma(\cdot)$ is the sigmoid function and $\gamma{=}10$ controls transition sharpness.

Intuitively, high-confidence classes receive $w_i \approx 0$, allowing full geometric correction by \methodname{}, whereas low-confidence polyphonic classes receive larger weights that preserve more of the original embedding geometry and reduce representation collapse under CCVD.
Importantly, CAR is integrated directly within the transductive calibration loop rather than applied as a post-processing heuristic. After each OPCA to CCVD iteration produces processed embeddings $\mathbf{E}^{\mathrm{PRISM}}$, CAR generates confidence-weighted targets $\tilde{\mathbf{E}}$, which ABR uses to compute the final affine correction matrix $\mathbf{W}=(\mathbf{E}_{\mathrm{aug}}^\top\mathbf{E}_{\mathrm{aug}}+\lambda\mathbf{I})^{-1}\mathbf{E}_{\mathrm{aug}}^\top\tilde{\mathbf{E}}$.

This incorporates polyphonic protection directly into the learned correction matrix $\mathbf{W}$ while preserving the same inference-time efficiency as the original PRISM framework. Empirically, CAR recovers \textit{street\_music} from $75.33\%$ to $83.49\%$. This introduces a necessary trade-off: soft interpolation attenuates the aggressive CCVD denoising on impulsive classes, causing partial regressions (e.g., \textit{air\_conditioner} drops $64.77\% \rightarrow 58.35\%$). Ultimately, CAR acts as a regularizer, trading a $0.48$ pp drop in peak overall accuracy for improved worst-case stability, while maintaining a strong $71.23\%$ overall accuracy ($+3.35$ pp above PCA++).

%%%%%%%%%%%%%%%%%%%%%%%%%%%%%%%%%%%%%%%%%%%%%%%%%%%%%%%%
\section{Conclusion}
\label{sec:conclusion}

We introduced \methodname{}, a gradient-free, source-free, and noise-prompt-free transductive TTA framework for Audio-Text Foundation Models under severe acoustic shift. Under the \textit{Affine Noise Hypothesis}, \methodname{} treats this shift as an approximately low-rank affine distortion and reverses it using OPCA, CCVD, and ABR, compiled into one static projection matrix. This separates batch-dependent calibration from sample-wise, batch-independent inference, enabling sub-millisecond adaptation without gradients or trainable parameters. Across additive-noise benchmarks, \methodname{} consistently outperforms strong blind baselines and oracle-assisted ContextDA on US8K, while Confidence-Aware Regression mitigates the \textit{Polyphonic Trap} for broadband classes. These results establish embedding-space geometric correction as an effective, efficient alternative to gradient-based TTA for robust audio-text inference.
\begin{acks}
Rini acknowledges financial support from the Department of Science and Technology, Government of India, through the WISE Post-Doctoral Fellowship Program (Ref.~DST/WISE-PDF/ET33/2023), which supported this work.
\end{acks}
\section*{GenAI Usage Disclosure}

Generative AI tools (e.g., large language models) were used to assist
with grammar checking and minor phrasing improvements in the manuscript.
All scientific claims, experimental designs, results, analyses, figures,
and theoretical derivations are entirely the work of the human authors.
No AI-generated content was used for data, methodology, or the
substantive intellectual contributions of this work.

\bibliographystyle{ACM-Reference-Format}
\bibliography{references}

\end{document}